\documentclass[10pt]{iopart}
\usepackage{epsfig,cite}
\newcommand{\beq}{\begin{equation}}
\newcommand{\eeq}{\end{equation}}

\newcommand{\beqar}{\begin{eqnarray}}
\newcommand{\eeqar}{\end{eqnarray}}

\begin{document}

\title[The nuclear equation of state probed by 
$K^+$ production in heavy ion collisions]{
The nuclear equation of state probed by 
$K^+$ production in heavy ion collisions}
\author{
C~Fuchs\dag, Amand~Faessler\dag, S El-Basaouny\dag, 
K Shekhter\dag, E~E~Zabrodin\dag\S, Y~M Zheng\ddag}
\address{\dag\
Institut f\"ur Theoretische Physik, Universit\"at
    T\"ubingen, T\"ubingen, Germany}
\address{\ddag\
China Institute of Atomic Energy, Beijing 102413, China}
\address{\S\
Institute for Nuclear Physics, Moscow State University, Moscow, 
    Russia}
\begin{abstract}
The dependence of  $K^+$ production on the nuclear equation of state 
is investigated in heavy ion collisions. 
An increase of the excitation function of $K^+$ 
multiplicities obtained in heavy ($Au+Au$) over light ($C+C$) systems 
when going far below threshold which has been observed by the 
KaoS Collaboration strongly favours a soft equation of state. This 
observation holds despite of the influence of an in-medium kaon 
potential predicted by effective chiral models which is necessary 
to reproduce the experimental $K^+$ yields. Phase space effects 
are discussed with respect to the $K^+$ excitation function.
\end{abstract}

\vspace{-0.5cm}
\section{Introduction and model}
\vspace{-0.1cm}
The original motivation to study the kaon production 
in heavy ion reactions at intermediate energies, namely to extract 
information on the nuclear equation of state (EOS) at high densities is 
a matter of current debate. Already in the first theoretical 
investigations by transport models it was noticed that the $K^+$ 
yield reacts sensitive on the nuclear equation of state 
\cite{AiKo85,qmd93,li95b,baoli94}. The yields were 
found to be about a factor 2--3 larger when a 
soft EOS was applied compared 
to a hard EOS. At that time the available data \cite{kaos94} 
already favoured a soft equation of state. However, calculations 
as well as the experimental data were still burdened with 
large uncertainties. 

In \cite{fuchs01} we studied the question if in the meantime 
decisive information on the nuclear EOS can be extracted from subthreshold 
kaon production in heavy ion collisions. There are several reasons 
why it appears worthwhile to do this: Firstly, there has been 
significant progress in the recent years towards a more precise 
determination of the elementary kaon production cross sections 
\cite{sibirtsev95,tsushima99}, based also on new data points form 
the COSY-11 for the reactions 
$pp \longrightarrow p K^+ X$ very close to threshold \cite{cosy11}. 
Secondly, the KaoS Collaboration has performed systematic 
measurements of the $ K^+$ production far below threshold in 
heavy ($Au+Au$) and light ($C+C$) systems \cite{sturm00}. Looking at the 
ratios built from heavy and light systems possible uncertainties which 
might still exist in the theoretical calculations should cancel out 
to a large extent which allows to draw reliable conclusions. 
Furthermore, far below threshold the kaon production is a highly 
collective process and a particular sensitivity to the compression of the 
participant matter is expected. 

The present investigations are based on the 
Quantum Molecular Dynamics (QMD) transport model \cite{ai91}. 
For the nuclear EOS we adopt soft and hard Skyrme forces 
corresponding to a compression modulus of 
K=200 MeV and 380 MeV, respectively, and with a momentum dependence 
adjusted to the empirical optical nucleon-nucleus potential 
\cite{ai91}. The saturation 
point of nuclear matter is thereby fixed at $E_B = -16$ MeV and 
$\rho_{\rm sat}= 0.17~{\rm fm}^{-3}$ \cite{ai91}. 
The calculations include $\Delta (1232)$ and $N^*(1440)$ resonances. 
The QMD approach with Skyrme interactions is well tested, 
contains a controlled momentum dependence and provides 
a reliable description of the reaction dynamics in the SIS 
energy range, expressed e.g. by collective nucleon flow observables as well 
as particle production. In contrast to AGS energies where 
the creation of resonance matter may lead to an effective 
softening of the EOS, baryonic resonances with masses 
above the $N^*(1440)$ can safely be neglected for the reaction 
dynamics at SIS energies \cite{hofmann95}.

We further consider the influence of an 
in-medium kaon potential based on 
effective chiral models \cite{kapla86,weise93,li95,brown962}. 
The $K^+$ mean field consists of a repulsive vector part 
$V_\mu = 3/8 f^{*2}_{\pi} j_\mu$ and an attractive scalar part 
$\Sigma_S = m_{{\rm K}} -  m_{{\rm K}}^* = m_{{\rm K}} - 
\sqrt{ m_{{\rm K}}^2 - \Sigma_{\mathrm{KN}}/f_{\pi}^2\rho_S 
     + V_\mu V^\mu }$. Here $j_\mu$ is the baryon vector current 
and $\rho_S$ the scalar baryon density and  
$\Sigma_{\mathrm{KN}} = 450$ MeV. Following \cite{brown962} in 
the vector field the pion decay constant in the medium 
$f^{*2}_\pi = 0.6 f^{2}_\pi$ is used. However, 
the enhancement of the scalar part using $f^{*2}_\pi$ is compensated 
by higher order contributions in the chiral expansion \cite{brown962}, 
and therefore here the bare value is used, i.e. 
$\Sigma_{\mathrm{KN}}\rho_S /f^{2}_\pi$.   
Compared to other chiral approaches \cite{weise93,li95} the 
resulting kaon dispersion relation shows a relatively strong 
density dependence. The increase of the 
in-medium $K^+$ mass ${\tilde m}_{\mathrm K}$, 
Eq. (\ref{effmass}), with this parameterisation 
is still consistent with the empirical 
knowledge of kaon-nucleus scattering and allows to explore 
in-medium effects on the production 
mechanism arising from zero temperature kaon potentials. 
For the kaon production via pion absorption $\pi B\longrightarrow YK^+ $ 
the elementary cross section of \cite{tuebingen1} are used. 
For the $N N \longrightarrow BYK^+ $ channels we apply 
the cross sections of Ref. \cite{sibirtsev95} which give a good fit 
to the COSY-data close to threshold. For the case of 
$N \Delta \longrightarrow BYK^+ $ and 
$\Delta \Delta \longrightarrow BYK^+ $ reactions experimental data 
are rare. Thus we rely on the model calculation of ref. \cite{tsushima99}. 
In the case that 
a $N^*$ resonance is involved in the reaction we used the same 
cross section as for nucleons. 
In the presence of scalar and vector fields the kaon optical potential 
in nuclear matter has the same structure as the corresponding 
Schroedinger equivalent optical potential for nucleons 
\beq
U_{\rm opt}(\rho ,{\bf k}) 
=  -\Sigma_S + \frac{1}{m_{\mathrm K}} k_{\mu} V^{\mu}  
+ \frac{\Sigma_S^2 - V_{\mu}^2}{2m_{\mathrm K}}  ~.
\label{uopt}
\end{equation}
and leads to a shift of the thresholds conditions inside 
the medium. To fulfil 
energy-momentum conservation the optical potential is absorbed 
into an newly defined effective mass 
\beq
{\tilde m}_{\mathrm K} (\rho ,{\bf k}) 
= \sqrt{ m_{\mathrm K}^2 + 2m_{\mathrm K} U_{\rm opt}(\rho ,{\bf k}) }
\label{effmass}
\eeq
which is a Lorentz scalar and sets the canonical momenta on the mass-shell 
$0=  k_{\mu}^{2} - {\tilde m}_{\mathrm K}^{2}$. 
Thus, e.g., the threshold condition for $K^+$ production in baryon induced 
reactions reads 
$\sqrt{s} \ge {\tilde m}_B + {\tilde m}_Y + {\tilde m}_K$ 
with $\sqrt{s}$ the centre--of--mass energy of the colliding baryons. 
For a consistent treatment of the thresholds the scalar 
and vector baryon mean fields entering into eq. 
(\ref{effmass}) are determined from two 
versions of the non-linear 
Walecka model with K=200/380 MeV, respectively \cite{li95b}. The hyperon 
field is thereby scaled by 2/3 which yields also a 
good description of the $\Lambda$ flow \cite{wang99b}. Since the 
parameterisations chosen for the non-linear Walecka model yield 
the same EOS as the Skyrme ones, the overall energy is conserved. 
The kaon production is treated perturbatively and does generally 
not affect the reaction dynamics \cite{fang94}.
\section{EOS dependence of $K^+$ production}
The $K^+$ excitation function for $Au+Au$ and $C+C$ 
reactions starting from 0.8 A$\cdot$GeV which is far below 
threshold ($E_{thr}=$ 1.58 GeV) has been measured by the KaoS 
Collaboration \cite{sturm00,kaos99}. In \cite{fuchs01} we 
calculated this excitation function for a soft/hard EOS 
including the in-medium kaon potential. 
For both systems the agreement with the 
KaoS data \cite{sturm00} is very good when a soft EOS is used. 
In the large system there was 
a visible EOS effect which is absent in the light system. 
The inclusion of the repulsive in-medium $K^+$ potential is 
thereby essential to reproduce the data \cite{kaos99}.  
Already in the light system the $K^+$ yield is reduced by about 
$50\%$. To extract more clear information on the nuclear EOS,  
in Fig. 1 we considered the ratio $R$ of the 
kaon multiplicities obtained in $Au+Au$ over $C+C$ 
reactions, normalised to the corresponding mass numbers.
The kaon potential is included since without the in-medium potential 
one is not able to reproduce the experimental $K^+$ yields 
\cite{fuchs01,hartnack01b}. 
\begin{figure}[h]
\unitlength1cm
\begin{picture}(8.,8.2)
\put(1.5,0){\makebox{\epsfig{file=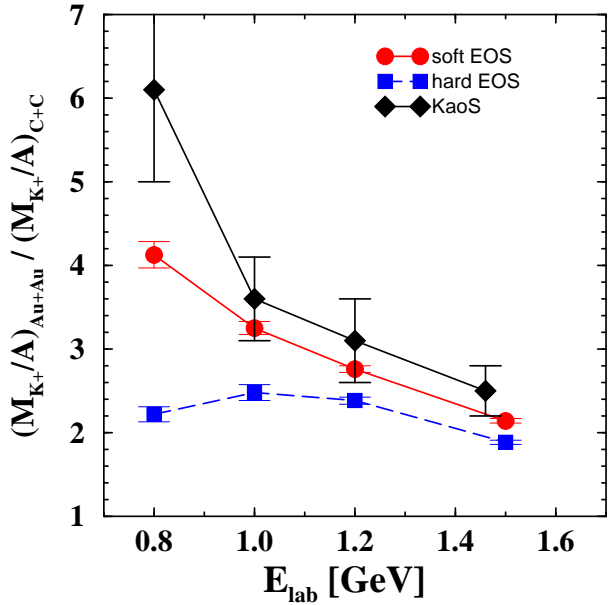,width=8.0cm}}}
\end{picture}
\caption{Excitation function of the ratio $R$ of $K^+$ 
multiplicities obtained in inclusive $Au+Au$ over $C+C$ 
reactions. The calculations are performed with 
in-medium kaon potential and using a hard/soft nuclear EOS and 
compared to the data from the KaoS Collaboration \protect\cite{sturm00}. 
}
\label{Fig3}
\end{figure}
The calculations are performed under 
minimal bias conditions with $b_{{\rm max}}=11$ fm 
for $Au+Au$ and $b_{{\rm max}}=5$ fm for $C+C$ and normalised to the 
experimental reaction cross sections \cite{sturm00,kaos99}. Both 
calculations show an increase of $R$ with decreasing incident energy 
down to 1.0 A$\cdot$GeV. However, this increase is much less 
pronounced when the stiff EOS is employed. 
In the latter  case $R$ even decreases at 0.8 A$\cdot$GeV 
whereas the soft EOS leads to an unrelieved increase of $R$. 
At 1.5 A$\cdot$GeV which is already very close to threshold 
the differences between the two models become small. 
The strong increase of $R$ can be directly related to 
higher compressible nuclear matter. The comparison to the experimental 
data from KaoS \cite{sturm00} where the increase of $R$ is even more 
pronounced strongly favours a soft equation of state. 
We would like to mention that similar results were also 
obtaind by independent IQMD calculations 
\cite{hartnack01b,hartnack01c}. These also include 
an in-medium kaon potential derived in 
relativistic mean field theory (RMF) \cite{schaffner97} 
which is somewhat less repulsive than that one 
used in our calculations. For the soft 
EOS the IQMD calculations almost coincide with the present 
results \cite{fuchs01}. For the hard EOS there exists still deviations 
concerning the slope of $R$ going far below threshold. This could be due 
to the different in-medium potentials and is an open question which 
has to be resolved by future investigations. However, the 
compared to earlier works \cite{hartnack01b} improved statistics 
has now led to a relatively good overall agreement of the two 
sets of transport calculations.  
\begin{figure}[h]
\unitlength1cm
\begin{picture}(10.,8)
\put(1.5,0){\makebox{\epsfig{file=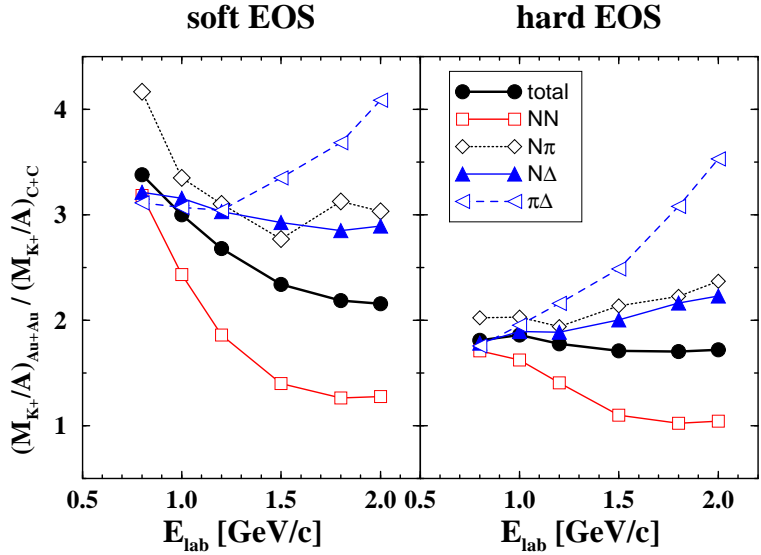,width=10.0cm}}}
\end{picture}
\caption{Dependence of the excitation function of $R$ on the 
various $K^+$ production channels. Central (b=0 fm) $Au+Au$ and $C+C$ 
reactions are considered. The calculations are performed with 
in-medium kaon potential.
}
\end{figure}

The dependence of $R$ on the various production channels is shown in Fig.2. 
There the ratios $R_i$ are built separately for the production 
channels with initial states $ i= NN, \pi N, N\Delta, \pi\Delta, 
\Delta\Delta$. Here one has to keep in mind that 
the $NN$ and $ \pi N$ cross sections 
are quite well under control whereas the $N\Delta~, \pi\Delta$ channels 
are experimentally unknown. Thus one has to rely on model predictions 
\cite{tsushima99}. However, the shape of $R$ is not strongly influenced 
by these two channels which are the most insecure ones. 
The excitation function for 
the $N\Delta$ contribution varies only little as a function of energy 
and is similar using the different EOSs. The contribution of the 
$\pi\Delta$ channel is decreasing for 
both, a hard and a soft EOS. The shape of $R$ is to most extent 
determined by the $NN$ and $ \pi N$ contributions. In our calculations 
the latter channel is responsible for the decrease of $R$ very far 
below threshold when the hard EOS is applied. Since we consider ratios 
theoretical uncertainties in the knowledge of elementary cross sections 
cancel out in first order anyway, which makes the conclusions more 
reliable. Also the Nantes group \cite{hartnack01c} 
reported that varying e.g. the $N\Delta$ cross section 
by a factor of two does hardly affect the final shape of $R$.   
\vspace{-0.1cm}
\section{Phase space for $K^+$ production}
\vspace{-0.1cm}
To obtain a quantitative picture of the explored density 
effects in Fig. 3 the baryon densities are shown at 
which the kaons are created. The energy is chosen most below 
threshold, i.e. at 0.8 A$\cdot$GeV and only central collisions 
are considered where the effects are maximal. $dM_{K^+}/d\rho$ 
is defined as 
\beq
dM_{K^+}/d\rho = \sum_{i}^{N_{K^+}} 
\frac{d P_i }{ d\rho_B ({\bf x}_i, t_i)}
\eeq 
where $\rho_B$ is the baryon density at which the kaon $i$ was created 
and  $P_i$ is the corresponding production probability. For 
the comparison of the two systems the curves are normalised to the 
corresponding mass numbers. 
\begin{figure}[h]
\unitlength1cm
\begin{picture}(8.,8.2)
\put(1.5,0){\makebox{\epsfig{file=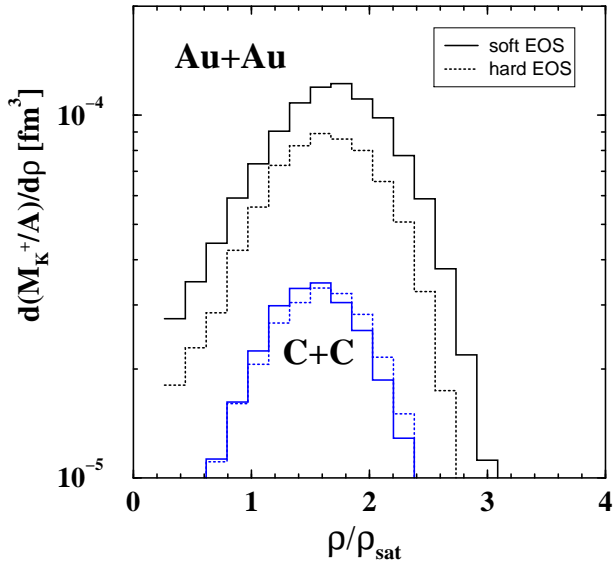,width=8.0cm}}}
\end{picture}
\caption{Kaon multiplicities (normalised to the mass numbers of the 
colliding nuclei) as a function of the baryon density 
at the space-time coordinates 
where the $K^+$ mesons have been created. Central (b=0 fm) $Au+Au$ and $C+C$ 
reactions at 0.8 A$\cdot$GeV are considered. 
The calculations are performed with 
in-medium kaon potential and using a hard/soft nuclear EOS. 
}
\label{Fig4}
\end{figure}
Fig.3 illustrates several features: 
Only in the case of a soft EOS the mean densities at which kaons 
are created differ significantly for the two different reaction 
systems, i.e. $<\rho /\rho_{\rm sat} >$=1.46/1.40 for $C+C$  
and 1.47/1.57 for $Au+Au$ using  
the hard/soft EOS. Generally, in $C+C$ reactions densities above 
$2\rho_{\rm sat}$ are rarely reached whereas in $Au+Au$ the kaons are 
created at densities up to three times saturation density. 
Furthermore, for $C+C$ the density distributions are weakly 
dependent on the nuclear EOS. The situation changes 
completely in $Au+Au$. Here the densities profile 
shows a pronounced EOS dependence \cite{li95b}. 
Moreover, the excess of kaons obtained with the soft EOS 
originates almost exclusively from 
high density matter which demonstrates that compression effects 
are probed. 
\begin{figure}[h]
\unitlength1cm
\begin{picture}(8.,9.5)
\put(1.5,0){\makebox{\epsfig{file=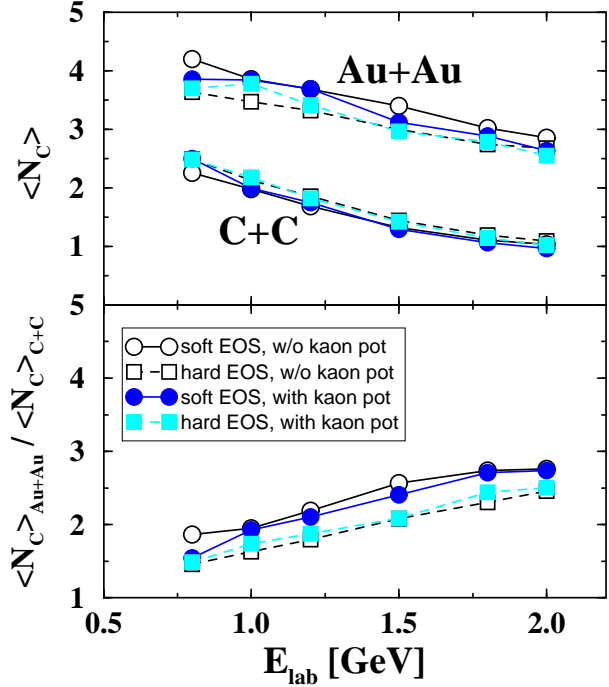,width=8.0cm}}}
\end{picture}
\caption{As a measure for the available phase space for $K^+$ 
production the mean number of collisions $<N_C >$ per particle 
which the hadrons ($N,\Delta,\pi$) did undergo before they produce 
a $K^+$ meson is considered. The upper panel shows $<N_C >$ in 
central $Au+Au$ and $C+C$ collisions. The lower panel shows the 
ratio of this quantity in $Au+Au$ over the same in $C+C$ 
reactions. The calculations are performed with/without 
in-medium kaon potential and using a hard/soft nuclear EOS.
}
\label{Fig5}
\end{figure}
Similar as the density shown before a quantitative measure 
for the collectivity probed by the  $K^+$ production and for  
phase space effects is shown in Fig.4. There the 
average number of collisions for those 
hadrons ($N, \Delta, \pi$) which were involved in the $K^+$ production 
are displayed. Again only central collisions are considered where the 
effects are maximal. $<N_C >$ is defined as 
\beq
<N_C > = \sum_{i}^{N_{K^+}} \frac{1}{2} (N_{C_{1}}^i + N_{C_{2}}^i) 
P_i / \sum_{i}^{N_{K^+}} P_i
\eeq 
with $ N_{C}^{i}$ being the number of collisions which particles  
($1,2$) experienced before they produced kaon $i$, and $P_i$ is the 
corresponding production probability. It is seen 
that in average the particles 
undergo about twice as much relevant collisions in the heavy 
compared to the light system. Furthermore, the collectivity, i.e. 
the accumulation of energy by multiple scattering, increases 
with decreasing incident energy. Thus one can conclude that the 
increase of $R$ is not due to a trivial phase space effect, namely 
the fact that far below threshold the $C+C$ system is  
simply too small to provide enough collectivity for the 
kaon production. If such a scenario - which could model independently 
as well explain the rise of $R$ seen in the KaoS data - 
would be true, $<N_C >$ would have to saturate for $C+C$ 
collisions at low energies. One can expect such a saturation 
from the number of binary collisions at even 
lower incident energies but here this is obviously not yet the case. 
Moreover, building also here the ratio (lower panel of Fig.4) it seems that 
the relative enhancement of available phase space for $K^+$ 
production in the large system is decreasing at low energies. 
This demonstrates that $K^+$ production 
far below threshold always requires a certain amount of collectivity 
which can be provided also in a very small colliding system, though 
such processes are rare. There is, however, no sharp limit 
were such collision histories become impossible. 
Thus trivial phase space effects can be excluded 
for an explanation of the increase of $R$. In \cite{sturm00} 
a similar argument was based on the measurement 
of high energy pions which can test the phase space available for 
particle production.
\vspace{-0.1cm}
\section{$K^+$ flow}
\vspace{-0.1cm}
\begin{figure}[h]
\unitlength1cm
\begin{picture}(8.,8.)
\put(1.5,0){\makebox{\epsfig{file=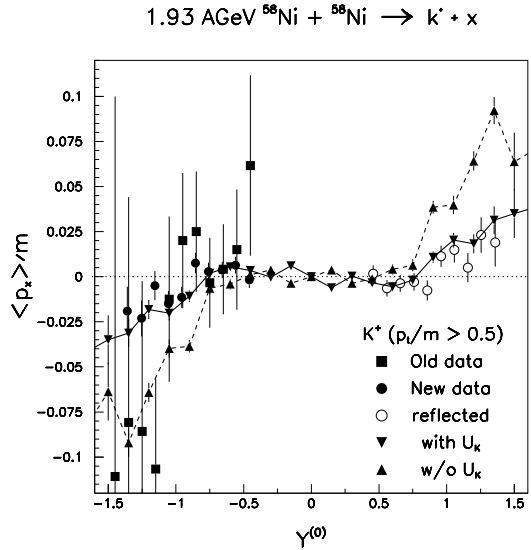,width=8.0cm}}}
\end{picture}
\caption{Average in-plane transverse $K^{+}$ flow 
in 1.93 A.GeV $^{58}$Ni + $^{58}$Ni reactions.
The full squares (circles) represent old (new) experimental data from 
FOPI \protect\cite{fopi95} (\protect\cite{fopi99}).  
The full down (up) triangles denote the calculated 
results with (without) kaon potential in the
nuclear medium. 
}
\label{Fig6}
\end{figure}
Finally in Fig.5 the transverse $K^+$ flow in $Ni+Ni$ reactions 
at 1.93 A.GeV is considered in order to obtain a conclusive picture 
of the consistency of in-medium effects with data \cite{zheng01}. 
As proposed in \cite{fuchs98} the $K^+$ in-medium potential is treated in its 
full covariant form, i.e. including the Lorentz force contribution 
which arises from the existence of the vector field. As discussed 
in \cite{fuchs98} the Lorentz force leads to a cancellation of 
the anti-flow \cite{li95,cassing99,brat98} due to the 
repulsive time-like  part of the vector potential. Thus there is 
no difference of the $K^+$ flow for the calculations with and without 
in-medium effects around mid-rapidity. However, at target and 
spectator rapidities the reduced error bars of more recent FOPI 
data \cite{fopi99} allow a distinction between the two scenarios. 
Both calculations show flow there but the flow signal is much weaker 
using the in-medium potential. Only the latter case is consistent with 
the data. The necessity to include the in-medium effects, on the other 
hand side, is consistent with knowledge from other dynamical observables 
\cite{cassing99,wang97}. 
\vspace{-0.1cm}
\section{Summary}
\vspace{-0.1cm}
To summarise, we find that at incident energies far below 
the free threshold $K^+$ production is a suitable tool to study the 
dependence on the nuclear equation of state. Using a light system 
as reference frame there is a visible sensitivity on the EOS 
when ratios of heavy ($Au+Au$) over light ($C+C$) systems are considered. 
Transport calculations indicate that the $K^+$ production gets 
hardly affected by compressional effects in $C+C$ but is 
highly sensitive to the high density matter 
($1\le \rho /\rho_{\rm sat} \le 3$) created in $Au+Au$ reactions. 
Results for the  $K^+$ excitation function in $Au+Au$ over 
$C+C$ reactions as measured by the KaoS Collaboration, strongly 
support the scenario with a soft EOS. This statement is also 
valid when an enhancement of the in-medium kaon mass as predicted 
by chiral models is taken into account. Since the explanation 
of the total $K^+$ yields and the  $K^+$ flow requires 
the presence of in-medium effects a consistent picture 
for the $K^+$ dynamics is obtained.
\\

The authors would like to acknowledge valuable discussions 
with J. Aichelin, Ch. Hartnack, H. Oeschler, P. Senger and C. Sturm. 

\end{document}